\documentclass[10pt,aps,ams,nofootinbib]{revtex4}
\usepackage{graphicx}
\usepackage{amsmath}
\usepackage{amsfonts}
\newcommand{\be}{\begin{equation}}
\newcommand{\ee}{\end{equation}}
\newcommand{\bea}{\begin{eqnarray}}
\newcommand{\eea}{\end{eqnarray}}
\newcommand{\bean}{\begin{eqnarray*}}
\newcommand{\eean}{\end{eqnarray*}}
\usepackage{latexsym}
\usepackage[T1]{fontenc}
\begin{document}
%\DocInput{draftwatermark.dtx}
\title{Two point function for a simple general relativistic quantum model}

%\date{}
\author{Daniele Colosi}
\email[Email address: ]{colosi@matmor.unam.mx}

\affiliation{Instituto de Matem\'aticas, UNAM, Campus Morelia\\ C.P. 58190, Morelia, Michoac\'an, Mexico}

\begin{abstract}
We study the quantum theory of a simple general relativistic quantum model of two coupled harmonic oscillators and compute the two-point function following a proposal first introduced in the context of loop quantum gravity.
\end{abstract}
\maketitle

\section{Introduction}

The quantum dynamics of a general relativistic system is not governed by a genuine Hamiltonian. Instead the evolution is fixed by the first class constraints of the theory, among which is the Hamiltonian constraint, and therefore evolution does not refer to an external (classical) time parameter - as is the case in standard quantum mechanics -  but corresponds to the unfolding of gauge transformations. The observables of the theory are gauge invariant quantities represented by operators commuting with the first class constraints and in particular with the Hamiltonian constraint and therefore are all constants of motion. This is at the heart of the so called "problem of time" for general relativistic systems: How can we reconcile a timeless quantum evolution with the experimental evidence of the flow of time? A related question is how can we extract from the quantum system a classical limit where the dynamical evolution is described with respect to a parameter playing the role of time.

In order to study such issues we examine a simple model of two harmonic oscillators characterized by an hamiltonian constraint. Such a model was first considered by Rovelli in \cite{Rovelli90} and its quantum dynamics was examined in details in \cite{pendolo}. In particular in this paper we compute and study the two-point function of this model, following a proposal introduced in \cite{Rovelli2005} to compute the graviton propagator in loop quantum gravity. In Sec. II we describe the classical and quantum dynamics of the system of two harmonic oscillators with an Hamiltonian constraint, and we give the expression of the projector operator from the kinematical to the physical Hilbert space of the quantum theory. Coherent states are introduced in Sec. III. They will play a key role in the following. Finally we compute explicitly in Sec. IV the two-point function of the model and comment some of its properties.

\section{Double harmonic oscillators}

Let us consider a classical system of two harmonic oscillators (of the same frequency and the same mass equal to 1) with positions $q_a,q_b$ and momenta $p_a,p_b$. The dynamics is governed by the hamiltonian constraint
\bea
H(q_a,q_b,p_a,p_b)= \frac{1}{2} \left(q_a^2+q_b^2+p_a^2+p_b^2 \right)-M \, \sim 0,
\label{H0}
\eea
where $M$ is a real positive constant characterizing the system. The extended configuration space is ${\cal C}= \mathbb R^2$ coordinatized by the observables quantities $q_a$ and $q_b$, called \textit{partial observables} \footnote{For the notion of partial observables as well as for a complete discussion on general relativistic quantum mechanics we refer to \cite{book}}. The function $H$ is defined on the cotangent space $T^*{\cal C}= \mathbb R^4$.

The action of the system is
\bea
S= \int d \tau \left(p_a \dot{q}_a+p_b \dot{q}_b -N \frac{1}{2}(q_a^2+q_b^2+p_a^2+p_b^2 -2M)\right),
\eea
where a dot denotes the derivative with respect to $\tau$ and $N$ is a lagrangian multiplier. The variation of the action with respect to the variables $q_a(\tau),p_a(\tau),q_b(\tau),p_b(\tau)$ and $N(\tau)$ gives the canonical equations of motion and the constraint equation (\ref{H0}). The action is invariant under reparametrization of the parameter $\tau$ labeling the motion.

The solution to the equations of motion can be parametrized by this non-physical parameter $\tau$
\bea
q_a(\tau) &=& A \, \sin(\tau+ \phi_a),
\label{EOM1}\\
q_b(\tau) &=& B \, \sin (\tau+ \phi_b).
\label{EOM2}
\eea
This motion is an ellipse of radii $A$ and $B$ and inclination $\Delta \phi= \phi_a-\phi_b$ in the extended configuration space ${\cal C}$, the plane $(q_a,q_b)$, with $(A,B) \in \left[ 0, \sqrt{2M}\right]$ and $(\phi, \phi') \in \left[0, 2 \pi \right]$. The values of the constants $A,B$ are fixed by the constraint (\ref{H0}) to satisfy $A^2+B^2=2M$. Due to the reparametrization invariance, the parameter $\tau$ is not connected with observability and the physical predictions of the system regard relation between the partial observables $a$ and $b$, not their dependence on $\tau$. A gauge transformation changes the phases $\phi_a$ and $\phi_b$ but not $\Delta \phi$. Therefore the physical phase space is coordinatized by the two quantities $A$ and $\Delta \phi$ which are constant along the motions, and identify the ellipse.

The canonical quantization of the system is straightforward: The partial observables $q_a,q_b$ and $p_a,p_b$ are promote to self-adjoint operators acting on the kinematical Hilbert space ${\cal K}= L_2 [\mathbb R^2, dq_a dq_b]$ as multiplicative operators and derivative operators respectively, and satisfying the usual commutation relations. The physical Hilbert space $\cal H$ of the theory is the space of solutions of the equation defining the dynamics, i.e. the Wheeler-DeWitt equation:
\bea
\left( - \frac{\partial^2}{\partial q_a^2}- \frac{\partial^2}{\partial q_b^2}+q_a^2+q_b^2-2M\right) \psi(q_a,q_b)=0,
\label{WDW}
\eea
where we have take $\hbar=1$. We introduce the creation and annihilation operators for the two oscillators, with the usual commutators $[a,a^{\dagger}]=[b,b^{\dagger}]=1$, $[a,b]=[a,b^{\dagger}]=0$, and the energy eigenstates $|n_a,n_b\rangle$ where $n_a$ and $n_b$ are the energy levels of the oscillators $q_a$ and $q_b$ respectively. The physical Hilbert space is spanned by the vectors satisfying the equation $M=n_a+n_b+1$. It is convenient to introduce the quantum number $m = \frac{1}{2}(n_a-n_b)$ that runs from $m=-j$ to $m=j$, with $j=(M-1)/2$. So that $\cal H$ is spanned by the $(2j+1)$ states $|m\rangle \equiv |j-m,j+m \rangle$. The representation of the states in the space of coordinates is
\bea
\psi_m(a,b)= \left(2^{2j}\pi (j+m)!(j-m)!\right)^{-1/2} H_{j+m}(a)H_{j-m}(b)e^{-\frac{a^2+b^2}{2}},
\eea
where $H_n(q)$ id the $n$-th Hermite polynomial. The dynamics can be obtained from an orthogonal projection operator $P$ from the kinematical Hilbert space $\cal K$ into the physical Hilbert space $\cal H$, $P: \cal K \rightarrow \cal H$, defined by
\bea
P= \int d\tau e^{-i \tau  H} = \sum_{m=-j}^j |m \rangle \langle m|,
\label{P}
\eea
and the propagator $K$ of the model is the integral kernel of $P$. The properties and the classical limit of the propagator $K$ have been studied in \cite{pendolo}.

\section{Coherent states}

The coherent states in the kinematical Hilbert space $\cal K$ can be written in terms of the energy eigenstates 
\bea
| \alpha,\beta \rangle = e^{-|\alpha|^2/2-|\beta|^2/2} \sum_{n_a,n_b}^{\infty} \frac{\alpha^{n_a}\beta^{n_b}}{\sqrt{n_a!n_b!}} |n_a,n_b\rangle,
\eea
where the parameters $\alpha,\beta$ are functions of the classical positions $(q_a,q_b)$ and the classical momenta $(p_a,p_b)$: $\alpha= q_a+i p_a$ and $\beta = q_b+ip_b$. To obtain the coherent states in the physical Hilbert space we use the projector $P$:
\bea
| \alpha,\beta \rangle_{phys} &=& P| \alpha,\beta \rangle \nonumber\\
&=& e^{-|\alpha|^2/2-|\beta|^2/2} \sum_{n_a+n_b+1=M} \frac{\alpha^{n_a}\beta^{n_b}}{\sqrt{n_a!n_b!}} |n_a,n_b\rangle\nonumber\\
&=& e^{-|\alpha|^2/2-|\beta|^2/2} \sum_{m=-j}^{j} \frac{\alpha^{j-m}\beta^{j+m}}{\sqrt{(j-m)!(j+m)!}} |m\rangle.
\eea
The wave function of this coherent state is peaked around the classical trajectory determined by the initial conditions $(q_a,q_b,p_a,p_b)$, namely around the ellipse with parameter $A$ and $\Delta \phi$ determined by $\alpha$ and $\beta$. The normalized coherent state is
\bea
| \alpha,\beta \rangle_{phys}^{norm} &=& \frac{1}{\sqrt{_{phys}\langle \alpha,\beta | \alpha,\beta \rangle_{phys}}} \,| \alpha,\beta \rangle_{phys} \nonumber\\
&=& \sum_{m=-j}^{j} \frac{\alpha^{j-m}\beta^{j+m}}{(|\alpha|^2+|\beta|^2)^j} \frac{\sqrt{(2j)!}}{\sqrt{(j-m)!(j+m)!}} |m\rangle.
\eea
Following the interpretation of general relativistic quantum mechanics given in \cite{book}, the scalar product between two coherent states represents a physical transition amplitude,
\bea
_{phys}^{norm}\langle \alpha',\beta' | \alpha,\beta \rangle_{phys}^{norm} = \frac{\left( \alpha'\alpha + \beta'\beta\right)^{2j}}{\left(|\alpha'|^2+|\beta'|^2 \right)^j \left(|\alpha|^2+|\beta|^2 \right)^j}.
\label{spc}
\eea
This product is in general different from zero, revealing the existence of some overlap between coherent states peaked around different classical solutions. However the modulus square of expression (\ref{spc}), having the interpretation of the transition probability form the state $| \alpha,\beta \rangle_{phys}^{norm}$ to the state $| \alpha',\beta' \rangle_{phys}^{norm}$, results to be maximal when $\alpha = \alpha'$ and $\beta = \beta'$. In \cite{Ashworth} Ashworth expresses the scalar product (\ref{spc}) in terms of the quantity $\xi=\alpha/\beta$, and shows that this scalar product, seen as a function of, say, $\xi'$ for $\xi$ fixed, is a Gaussian centered on $\xi$ with width equal to $(2j)^{-1}$. Therefore in the classical limit where $2j$ is large the transition between coherent states peaked around different classical trajectories is highly suppressed.

\section{Two-point function}

The two-point function for a single (non-relativistic) harmonic oscillator is given by
\bea
G(t_1,t_2)= \langle 0 | x(t_1)x(t_2)| 0 \rangle=\langle 0 | x\, e^{-iH(t_1-t_2)}\,x| 0 \rangle,
\label{G}
\eea
where $H$ is the hamiltonian and $\hbar=1$. The standard interpretation is that $G(t_1,t_2)$ represents the probability amplitude that a particle created at the instant $t_1$ is measured at the instant $t_2$. This definition of the two-point function appears to be hardly implementable for a general relativistic quantum mechanical system. The difficulties come from the absence of a vacuum state and a time parameter with respect to which evolution is defined for such system. How can we define, in a general relativistic quantum mechanical system, the distance between the two points appearing in (\ref{G})? How can we give an interpretation in terms of a particle propagating between the two points?

Recently a technique to compute the two-point function in the context of quantum gravity has been proposed in \cite{Rovelli2005}. We want to apply this proposal to our system of two harmonic oscillators that share with quantum gravity the basic property of not having any background time parameter and no state minimizing the energy, i.e. no vacuum state. The ingredients of the proposal are a boundary semiclassical state, the propagator of the theory, and operators acting on the boundary state. Following \cite{Rovelli2005} we take as boundary state a state living in the Hilbert space $\cal H^* \times \cal H$, in coordinate representation
\bea
\Psi(q_a,q_b,q_a',q_b')= \overline{\psi_1(q_a,q_b)} \, \psi_2(q_a',q_b'),
\label{coherentstate}
\eea
where $\psi_1(q_a,q_b)= \langle q_a,q_b | \alpha_1,\beta_1 \rangle_{phys}^{norm}$ and  $\psi_2(q_a,q_b)= \langle q_a,q_b | \alpha_2,\beta_2 \rangle_{phys}^{norm}$. The use of coherent state is the key idea in order to define the two points appearing in (\ref{G}), better the "temporal distance" between the two points. We choose the two coherent states $\psi_1$ and $\psi_2$ as peaked around the \textit{same} classical trajectory. Moreover the parameter $\alpha_1,\beta_1,\alpha_2,\beta_2$ will identify two particular points on this trajectory. For an ellipse given by the classical motion (\ref{EOM1},\ref{EOM2}), these parameters can be written as
\bea
\alpha_1 &=& A \, e^{-i(\tau_1+ \phi_a)},\label{a1}\\
\beta_1 &=& B \, e^{-i(\tau_1+ \phi_b)},\label{a2}\\
\alpha_2 &=& A \, e^{-i(\tau_2+ \phi_a)},\label{a3}\\
\beta_2 &=& B \, e^{-i(\tau_2+ \phi_b)}\label{a4}.
\eea
Therefore the distance between the two points will be measured by the difference of the phase factors, $\Delta \tau = \tau_1-\tau_2$. 

The next ingredient is easly obtained, it's the propagator of the model, namely the matrix elements of the projector (\ref{P}) on the physical Hilbert space, $K(q_a,q_b,q_a',q_b')= \langle q_a, q_b |P| q_a', q_b' \rangle$.  

Finally we need the operators acting on the boundary state. For that we simply take the product of the positions operators $q_a$ and $q_b$. We notice that these operators are not gauge invariant, because they do not commute with the hamiltonian constraint. In particular the product $q_a q_b$ contains a gauge invariant part, given in terms of the creation and annihilation operators by $ab^{\dagger}+a^{\dagger}b$, and a non gauge invariant part, $ab+a^{\dagger}b^{\dagger}$. The role of the propagator will be the one to select the gauge invariant part of $q_aq_b$.

We are now ready to compute the two-point function for our general relativistic quantum mechanical system following the proposal of \cite{Rovelli2005}. We indicate the two-point function as $G(\Delta \tau)$, where the argument of $G$ is connected with the distance of the two points. The expression is
\bea
G(\Delta \tau) = \int dq_adq_bdq_a'dq_b' \, \overline{\psi_1(q_a,q_b)} \,q_a q_b \, K(q_a,q_b,q_a',q_b')\, q_a'q_b'\, \psi_2(q_a',q_b'),
\label{gg}
\eea
or in the Dirac notation
\bea
G(\Delta \tau) =\, \, _{phys}^{norm} \langle \alpha_2,\beta_2| q_aq_b P q_a q_b |\alpha_1,\beta_1 \rangle_{phys}^{norm}.
\label{g1}
\eea
With simple algebra we obtain
\bea
G(\Delta \tau) = 2j(2j-1)\frac{(\overline{\alpha_2}\alpha_1+\overline{\beta_2}\beta_1)^{2j-2}}{(|\alpha_1|^2+|\beta_1|^2)^j(|\alpha_2|^2+|\beta_2|^2)^j} \left[(\overline{\beta_2}\alpha_1)^2+(\overline{\alpha_2}\beta_1)^2+2\overline{\alpha_2}\alpha_1\overline{\beta_2}\beta_1 + \frac{(\overline{\alpha_2}\alpha_1+\overline{\beta_2}\beta_1)^2}{2j-1}\right].
\eea
Substituting the parameters $\alpha$ and $\beta$ with their expression (\ref{a1}-\ref{a4}),
\bea
G(\Delta \tau) = \frac{2j(2j-1)}{M^2} \left( A^2B^2 \cos^2 \Delta \phi  + \frac{M^2}{2j-1}\right) e^{-i2j \Delta \tau}.
\label{g3}
\eea
In the semiclassical limit where $2j \sim M$ the two-point function takes the form
\bea
G(\Delta \tau) =  \left( A^2B^2 \cos^2 \Delta \phi  + M\right) e^{-iM \Delta \tau}.
\label{g2}
\eea
The result deserves various comments.
\begin{itemize}
	\item The two-point function is a gauge invariant quantity, and therefore can be seen as real physical prediction of the theory. This can be easly proved noting that a gauge transformation on the system transforms the complex quantities as $\alpha \rightarrow \alpha e^{i\theta}$ and $\beta \rightarrow \beta e^{i\theta}$. In the coherent states these gauge transformations appear as an overall phase in front of the vector, $|\alpha_1,\beta_1 \rangle_{phys}^{norm} \rightarrow  e^{i2j\theta}|\alpha_1,\beta_1 \rangle_{phys}^{norm}$. Hence the expression (\ref{g1}) of the two-point function is manifestly gauge invariant. In fact the resulting formula (\ref{g3}) (as well as the semiclassical limit (\ref{g2})) depends only on the difference of the phases of (\ref{a1}-\ref{a4}); a gauge transformation simply adds a constant quantity to these phases, and such quantity disappears in the difference.
	\item All the information about the distance of the two points is contained in the boundary state, namely in the coherent states determined by parameters (\ref{a1}-\ref{a4}). The presence of the propagator in the expression of the two-point function has the unique role to select the gauge invariant part of the product of the positions operators, $q_aq_b$. In contrast with the two-point function for the simple harmonic oscillator (\ref{G}) where the notion of the two points is coded in the propagator $\exp{-iH (t_1-t_2)}$, the action of the propagator in (\ref{gg}) only guarantees the gauge invariance of the full expression.
	\item The structure of the result (\ref{g3}) (or the semiclassical limit (\ref{g2})) has some analogies with the two-point function of a simple harmonic oscillator (\ref{G}). An easy computation gives (for convenience we explicitly write the mass $m$ and the frequency $\omega$ of the oscillator) 
\bea
G(t_1,t_2)= \langle 0 | x(t_1)x(t_2)| 0 \rangle= \frac{1}{2m\omega}e^{-i\frac{3}{2}\omega (t_1-t_2)}.
\label{2p}
\eea 
Hence the two-point function of a simple harmonic oscillator is composed by two factors, a constant term depending only on the characteristics of the system, and a complex exponential depending on the energy of the propagating particle and the time interval of the propagation. The two-point function for our system has a similar structure: Two factors, the first is a constant term function of the properties of the system (the quantity $M$) and of the particular classical trajectory selected by the boundary state (the parameters $A$ and $\Delta \phi$), and the second is a complex exponential of the "time interval" separating the two points.

It is worth noticing that the factor $\frac{3}{2} \omega$ appearing in the exponential of (\ref{2p}) represents the energy of all the particles involved in the process: In this case it is the energy of the first excited state of the quantum harmonic oscillator. In the very same way, the factor $2j$ in the exponential of the two-point function (\ref{g3}) represents the total energy of the particle taking part in the process: Such energy is fixed by the quantum hamiltonian constraint.
	\item Another property of the two-point function (\ref{gg}) is to be proportional to the scalar product of the coherent states that define the boundary state. In fact the scalar product of the physical states $|\alpha_1,\beta_1 \rangle_{phys}^{norm}$ and $|\alpha_2,\beta_2 \rangle_{phys}^{norm}$ is given by 
	\bea
	\, _{phys}^{norm} \langle \alpha_2,\beta_2|\alpha_1,\beta_1 \rangle_{phys}^{norm}= e^{-i2j \Delta \tau},
	\eea
	which is exactly the exponential appearing in (\ref{g3}).
\end{itemize}

\section{Summary}

We have computed the two-point function for the system of two quantum harmonic oscillators which dynamics is fixed by an Hamiltonian constraint. The computation was based on the proposal of \cite{Rovelli2005} for the two-point function in the context of loop quantum gravity. In our model a key ingredient has been the use of coherent states defined in the physical Hilbert space and peaked around classical solutions. The information concerning the two points of the correlation function is entirely contained in the boundary state build on the tensor product of two coherent states. The notion of time and in particular the notion of the "temporal distance" between the two points emerges from an appropriate choice of the boundary state, and is not related with the propagator of the system whose role is only to ensure the gauge invariance of the two-point function.

\end{document}